\pdfoutput=1
\documentclass[preprint,  prd, tightenlines, superscriptaddress]{revtex4-2}
\usepackage{graphicx}
\usepackage{xcolor}
\usepackage{amssymb}
\usepackage{url}
\usepackage{amsmath}
\usepackage{hyperref}
\usepackage{amsfonts}
\usepackage{xspace}
\usepackage{lineno}
\usepackage{ulem}

\renewcommand{\arraystretch}{1.15}
\input belle2sym

\usepackage{relsize}
\def\babar {\texorpdfstring{{\mbox{\slshape B\kern-0.1em{\smaller A}\kern-0.1em
    B\kern-0.1em{\smaller A\kern-0.2em R}}}\xspace}{BaBar}}

\let\klongID=\KLID
\def\CP{\textit{CP}\xspace}
\def\Nlayers {\ensuremath{N_\mathrm{layers}}\xspace}
\def\fpeak {\ensuremath{f_\mathrm{peak}}\xspace}

\def\Fig#1{Figure \ref{#1}}

\def\bjpsikz {\ensuremath{\Bz\to\jpsi\Kz}\xspace}
\def\bjpsikl {\ensuremath{\Bz\to\jpsi\KL}\xspace}
\def\bjpsiks {\ensuremath{\Bz\to\jpsi\KS}\xspace}
\def\bjpsimumukl {\ensuremath{\Bz\to\jpsi[\mumu] \KL}\xspace}
\def\bjpsieekl {\ensuremath{\Bz\to\jpsi[\ee] \KL}\xspace}
\def\deltaE {\ensuremath{\Delta E}\xspace}

\def\ifb {\ensuremath{\mathrm{fb}^{-1}}\xspace}
\def\iab {\ensuremath{\mathrm{ab}^{-1}}\xspace}
\def\smixed {\ensuremath{\BzBzb}\xspace}
\def\scharged {\ensuremath{\BpBm}\xspace}
\def\sccbar {\ensuremath{c\bar c}\xspace}
\def\sampleA {signal MC\xspace}
\def\sampleB {generic MC\xspace}
\def\Nsig {\ensuremath{N_\mathrm{sig}}\xspace}

\makeatletter
\newcommand{\thickhline}{%
    \noalign {\ifnum 0=`}\fi \hrule height 1pt
    \futurelet \reserved@a \@xhline
}
\makeatother

\begin{document}

\vspace*{-3\baselineskip}
\resizebox{!}{2.8cm}{\includegraphics{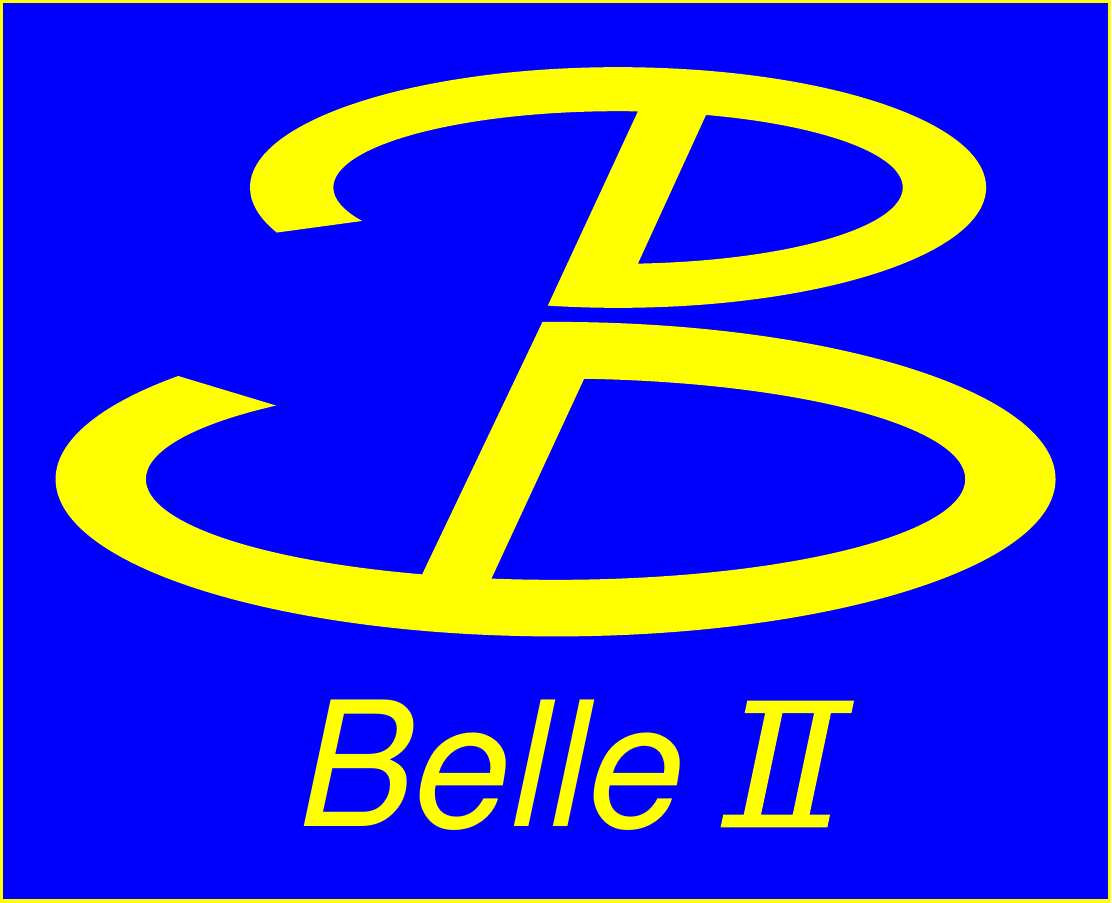}}
{\raggedright{\vbox{
      \hbox{{\hskip 6cm} BELLE2-CONF-PH-2021-009}
\hbox{{\hskip 6cm} \today}
\hbox{   }
\hbox{   }
}}
}
\title { \quad\\[0.5cm] \boldmath{Rediscovery of \bjpsikl at \belletwo}}

\newcommand{\instCPPM}{Aix Marseille Universit\'{e}, CNRS/IN2P3, CPPM, 13288 Marseille, France}
\newcommand{\instBeihang}{Beihang University, Beijing 100191, China}
\newcommand{\instBNL}{Brookhaven National Laboratory, Upton, New York 11973, U.S.A.}
\newcommand{\instBINP}{Budker Institute of Nuclear Physics SB RAS, Novosibirsk 630090, Russian Federation}
\newcommand{\instCMU}{Carnegie Mellon University, Pittsburgh, Pennsylvania 15213, U.S.A.}
\newcommand{\instCinvestavIPN}{Centro de Investigacion y de Estudios Avanzados del Instituto Politecnico Nacional, Mexico City 07360, Mexico}
\newcommand{\instPrague}{Faculty of Mathematics and Physics, Charles University, 121 16 Prague, Czech Republic}
\newcommand{\instChiangMai}{Chiang Mai University, Chiang Mai 50202, Thailand}
\newcommand{\instChiba}{Chiba University, Chiba 263-8522, Japan}
\newcommand{\instChonnam}{Chonnam National University, Gwangju 61186, South Korea}
\newcommand{\instConacyt}{Consejo Nacional de Ciencia y Tecnolog\'{\i}a, Mexico City 03940, Mexico}
\newcommand{\instDESY}{Deutsches Elektronen--Synchrotron, 22607 Hamburg, Germany}
\newcommand{\instDuke}{Duke University, Durham, North Carolina 27708, U.S.A.}
\newcommand{\instITAR}{ Duy Tan University, Hanoi 100000, Vietnam}
\newcommand{\instRomaENEA}{ENEA Casaccia, I-00123 Roma, Italy}
\newcommand{\instEri}{Earthquake Research Institute, University of Tokyo, Tokyo 113-0032, Japan}
\newcommand{\instJuelich}{Forschungszentrum J\"{u}lich, 52425 J\"{u}lich, Germany}
\newcommand{\instFuJen}{Department of Physics, Fu Jen Catholic University, Taipei 24205, Taiwan}
\newcommand{\instFudan}{Key Laboratory of Nuclear Physics and Ion-beam Application (MOE) and Institute of Modern Physics, Fudan University, Shanghai 200443, China}
\newcommand{\instGoettingen}{II. Physikalisches Institut, Georg-August-Universit\"{a}t G\"{o}ttingen, 37073 G\"{o}ttingen, Germany}
\newcommand{\instGifu}{Gifu University, Gifu 501-1193, Japan}
\newcommand{\instSOKENDAI}{The Graduate University for Advanced Studies (SOKENDAI), Hayama 240-0193, Japan}
\newcommand{\instGyeongsang}{Gyeongsang National University, Jinju 52828, South Korea}
\newcommand{\instHanyang}{Department of Physics and Institute of Natural Sciences, Hanyang University, Seoul 04763, South Korea}
\newcommand{\instKEK}{High Energy Accelerator Research Organization (KEK), Tsukuba 305-0801, Japan}
\newcommand{\instJPARC}{J-PARC Branch, KEK Theory Center, High Energy Accelerator Research Organization (KEK), Tsukuba 305-0801, Japan}
\newcommand{\instHSE}{Higher School of Economics (HSE), Moscow 101000, Russian Federation}
\newcommand{\instIISER}{Indian Institute of Science Education and Research Mohali, SAS Nagar, 140306, India}
\newcommand{\instIITBhubaneswar}{Indian Institute of Technology Bhubaneswar, Satya Nagar 751007, India}
\newcommand{\instIITGuwahati}{Indian Institute of Technology Guwahati, Assam 781039, India}
\newcommand{\instIITHyderabad}{Indian Institute of Technology Hyderabad, Telangana 502285, India}
\newcommand{\instIITMadras}{Indian Institute of Technology Madras, Chennai 600036, India}
\newcommand{\instIndiana}{Indiana University, Bloomington, Indiana 47408, U.S.A.}
\newcommand{\instIHEPRussia}{Institute for High Energy Physics, Protvino 142281, Russian Federation}
\newcommand{\instHEPHYVienna}{Institute of High Energy Physics, Vienna 1050, Austria}
\newcommand{\instHiroshima}{Hiroshima University, Higashi-Hiroshima, Hiroshima 739-8530, Japan}
\newcommand{\instIHEPChina}{Institute of High Energy Physics, Chinese Academy of Sciences, Beijing 100049, China}
\newcommand{\instIPP}{Institute of Particle Physics (Canada), Victoria, British Columbia V8W 2Y2, Canada}
\newcommand{\instIOP}{Institute of Physics, Vietnam Academy of Science and Technology (VAST), Hanoi, Vietnam}
\newcommand{\instIFIC}{Instituto de Fisica Corpuscular, Paterna 46980, Spain}
\newcommand{\instFrascati}{INFN Laboratori Nazionali di Frascati, I-00044 Frascati, Italy}
\newcommand{\instNapoliINFN}{INFN Sezione di Napoli, I-80126 Napoli, Italy}
\newcommand{\instPadovaINFN}{INFN Sezione di Padova, I-35131 Padova, Italy}
\newcommand{\instPerugiaINFN}{INFN Sezione di Perugia, I-06123 Perugia, Italy}
\newcommand{\instPisaINFN}{INFN Sezione di Pisa, I-56127 Pisa, Italy}
\newcommand{\instRomaINFN}{INFN Sezione di Roma, I-00185 Roma, Italy}
\newcommand{\instRomaTreINFN}{INFN Sezione di Roma Tre, I-00146 Roma, Italy}
\newcommand{\instTorinoINFN}{INFN Sezione di Torino, I-10125 Torino, Italy}
\newcommand{\instTriesteINFN}{INFN Sezione di Trieste, I-34127 Trieste, Italy}
\newcommand{\instJAEA}{Advanced Science Research Center, Japan Atomic Energy Agency, Naka 319-1195, Japan}
\newcommand{\instMainz}{Johannes Gutenberg-Universit\"{a}t Mainz, Institut f\"{u}r Kernphysik, D-55099 Mainz, Germany}
\newcommand{\instGiessen}{Justus-Liebig-Universit\"{a}t Gie\ss{}en, 35392 Gie\ss{}en, Germany}
\newcommand{\instKarlsruhe}{Institut f\"{u}r Experimentelle Teilchenphysik, Karlsruher Institut f\"{u}r Technologie, 76131 Karlsruhe, Germany}
\newcommand{\instISU}{Iowa State University, Ames, Iowa 50011, U.S.A.}
\newcommand{\instKitasato}{Kitasato University, Sagamihara 252-0373, Japan}
\newcommand{\instKISTI}{Korea Institute of Science and Technology Information, Daejeon 34141, South Korea}
\newcommand{\instKoreaUnivKU}{Korea University, Seoul 02841, South Korea}
\newcommand{\instKSU}{Kyoto Sangyo University, Kyoto 603-8555, Japan}
\newcommand{\instKyungpook}{Kyungpook National University, Daegu 41566, South Korea}
\newcommand{\instLPI}{P.N. Lebedev Physical Institute of the Russian Academy of Sciences, Moscow 119991, Russian Federation}
\newcommand{\instLNNU}{Liaoning Normal University, Dalian 116029, China}
\newcommand{\instLMU}{Ludwig Maximilians University, 80539 Munich, Germany}
\newcommand{\instLuther}{Luther College, Decorah, Iowa 52101, U.S.A.}
\newcommand{\instMNITJaipur}{Malaviya National Institute of Technology Jaipur, Jaipur 302017, India}
\newcommand{\instMPP}{Max-Planck-Institut f\"{u}r Physik, 80805 M\"{u}nchen, Germany}
\newcommand{\instMPGHLL}{Semiconductor Laboratory of the Max Planck Society, 81739 M\"{u}nchen, Germany}
\newcommand{\instMcGill}{McGill University, Montr\'{e}al, Qu\'{e}bec, H3A 2T8, Canada}
\newcommand{\instMEPhI}{Moscow Physical Engineering Institute, Moscow 115409, Russian Federation}
\newcommand{\instNagoya}{Graduate School of Science, Nagoya University, Nagoya 464-8602, Japan}
\newcommand{\instNagoyaIAR}{Institute for Advanced Research, Nagoya University, Nagoya 464-8602, Japan}
\newcommand{\instNagoyaKMI}{Kobayashi-Maskawa Institute, Nagoya University, Nagoya 464-8602, Japan}
\newcommand{\instNaraWu}{Nara Women's University, Nara 630-8506, Japan}
\newcommand{\instNTUTaiwan}{Department of Physics, National Taiwan University, Taipei 10617, Taiwan}
\newcommand{\instNUUTaiwan}{National United University, Miao Li 36003, Taiwan}
\newcommand{\instKrakow}{H. Niewodniczanski Institute of Nuclear Physics, Krakow 31-342, Poland}
\newcommand{\instNiigata}{Niigata University, Niigata 950-2181, Japan}
\newcommand{\instNSU}{Novosibirsk State University, Novosibirsk 630090, Russian Federation}
\newcommand{\instOkinawa}{Okinawa Institute of Science and Technology, Okinawa 904-0495, Japan}
\newcommand{\instOsakaCity}{Osaka City University, Osaka 558-8585, Japan}
\newcommand{\instRCNP}{Research Center for Nuclear Physics, Osaka University, Osaka 567-0047, Japan}
\newcommand{\instPNNL}{Pacific Northwest National Laboratory, Richland, Washington 99352, U.S.A.}
\newcommand{\instPanjab}{Panjab University, Chandigarh 160014, India}
\newcommand{\instPanjabPAU}{Punjab Agricultural University, Ludhiana 141004, India}
\newcommand{\instRIKENMSL}{Meson Science Laboratory, Cluster for Pioneering Research, RIKEN, Saitama 351-0198, Japan}
\newcommand{\instSeoul}{Seoul National University, Seoul 08826, South Korea}
\newcommand{\instSPU}{Showa Pharmaceutical University, Tokyo 194-8543, Japan}
\newcommand{\instSoochow}{Soochow University, Suzhou 215006, China}
\newcommand{\instSoongsil}{Soongsil University, Seoul 06978, South Korea}
\newcommand{\instLjubljanaJSI}{J. Stefan Institute, 1000 Ljubljana, Slovenia}
\newcommand{\instKyiv}{Taras Shevchenko National Univ. of Kiev, Kiev, Ukraine}
\newcommand{\instTata}{Tata Institute of Fundamental Research, Mumbai 400005, India}
\newcommand{\instTUM}{Department of Physics, Technische Universit\"{a}t M\"{u}nchen, 85748 Garching, Germany}
\newcommand{\instTelAviv}{Tel Aviv University, School of Physics and Astronomy, Tel Aviv, 69978, Israel}
\newcommand{\instToho}{Toho University, Funabashi 274-8510, Japan}
\newcommand{\instTohoku}{Department of Physics, Tohoku University, Sendai 980-8578, Japan}
\newcommand{\instTitech}{Tokyo Institute of Technology, Tokyo 152-8550, Japan}
\newcommand{\instTokyoMetropolitan}{Tokyo Metropolitan University, Tokyo 192-0397, Japan}
\newcommand{\instUAS}{Universidad Autonoma de Sinaloa, Sinaloa 80000, Mexico}
\newcommand{\instNapoliUNIV}{Dipartimento di Scienze Fisiche, Universit\`{a} di Napoli Federico II, I-80126 Napoli, Italy}
\newcommand{\instPadovaUNIV}{Dipartimento di Fisica e Astronomia, Universit\`{a} di Padova, I-35131 Padova, Italy}
\newcommand{\instPerugiaUNIV}{Dipartimento di Fisica, Universit\`{a} di Perugia, I-06123 Perugia, Italy}
\newcommand{\instPisaUNIV}{Dipartimento di Fisica, Universit\`{a} di Pisa, I-56127 Pisa, Italy}
\newcommand{\instRomaTreUNIV}{Dipartimento di Matematica e Fisica, Universit\`{a} di Roma Tre, I-00146 Roma, Italy}
\newcommand{\instTorinoUNIV}{Dipartimento di Fisica, Universit\`{a} di Torino, I-10125 Torino, Italy}
\newcommand{\instTriesteUNIV}{Dipartimento di Fisica, Universit\`{a} di Trieste, I-34127 Trieste, Italy}
\newcommand{\instMontreal}{Universit\'{e} de Montr\'{e}al, Physique des Particules, Montr\'{e}al, Qu\'{e}bec, H3C 3J7, Canada}
\newcommand{\instIJCLab}{Universit\'{e} Paris-Saclay, CNRS/IN2P3, IJCLab, 91405 Orsay, France}
\newcommand{\instIPHC}{Universit\'{e} de Strasbourg, CNRS, IPHC, UMR 7178, 67037 Strasbourg, France}
\newcommand{\instAdelaide}{Department of Physics, University of Adelaide, Adelaide, South Australia 5005, Australia}
\newcommand{\instBonn}{University of Bonn, 53115 Bonn, Germany}
\newcommand{\instUBC}{University of British Columbia, Vancouver, British Columbia, V6T 1Z1, Canada}
\newcommand{\instCincinnati}{University of Cincinnati, Cincinnati, Ohio 45221, U.S.A.}
\newcommand{\instFlorida}{University of Florida, Gainesville, Florida 32611, U.S.A.}
\newcommand{\instHawaii}{University of Hawaii, Honolulu, Hawaii 96822, U.S.A.}
\newcommand{\instHeidelberg}{University of Heidelberg, 68131 Mannheim, Germany}
\newcommand{\instLjubljanaUniLJ}{Faculty of Mathematics and Physics, University of Ljubljana, 1000 Ljubljana, Slovenia}
\newcommand{\instLouisville}{University of Louisville, Louisville, Kentucky 40292, U.S.A.}
\newcommand{\instMalaya}{National Centre for Particle Physics, University Malaya, 50603 Kuala Lumpur, Malaysia}
\newcommand{\instLjubljanaUM}{University of Maribor, 2000 Maribor, Slovenia}
\newcommand{\instMelbourne}{School of Physics, University of Melbourne, Victoria 3010, Australia}
\newcommand{\instMississippi}{University of Mississippi, University, Mississippi 38677, U.S.A.}
\newcommand{\instUOM}{University of Miyazaki, Miyazaki 889-2192, Japan}
\newcommand{\instPittsburgh}{University of Pittsburgh, Pittsburgh, Pennsylvania 15260, U.S.A.}
\newcommand{\instUSTC}{University of Science and Technology of China, Hefei 230026, China}
\newcommand{\instSAlabama}{University of South Alabama, Mobile, Alabama 36688, U.S.A.}
\newcommand{\instSCarolina}{University of South Carolina, Columbia, South Carolina 29208, U.S.A.}
\newcommand{\instSydney}{School of Physics, University of Sydney, New South Wales 2006, Australia}
\newcommand{\instUTokyo}{Department of Physics, University of Tokyo, Tokyo 113-0033, Japan}
\newcommand{\instIPMU}{Kavli Institute for the Physics and Mathematics of the Universe (WPI), University of Tokyo, Kashiwa 277-8583, Japan}
\newcommand{\instVictoria}{University of Victoria, Victoria, British Columbia, V8W 3P6, Canada}
\newcommand{\instVPI}{Virginia Polytechnic Institute and State University, Blacksburg, Virginia 24061, U.S.A.}
\newcommand{\instWayneState}{Wayne State University, Detroit, Michigan 48202, U.S.A.}
\newcommand{\instYamagata}{Yamagata University, Yamagata 990-8560, Japan}
\newcommand{\instYerevan}{Alikhanyan National Science Laboratory, Yerevan 0036, Armenia}
\newcommand{\instYonsei}{Yonsei University, Seoul 03722, South Korea}
\newcommand{\instZZU}{Zhengzhou University, Zhengzhou 450001, China}
\affiliation{\instCPPM}
\affiliation{\instBeihang}
\affiliation{\instBNL}
\affiliation{\instBINP}
\affiliation{\instCMU}
\affiliation{\instCinvestavIPN}
\affiliation{\instPrague}
\affiliation{\instChiangMai}
\affiliation{\instChiba}
\affiliation{\instChonnam}
\affiliation{\instConacyt}
\affiliation{\instDESY}
\affiliation{\instDuke}
\affiliation{\instITAR}
\affiliation{\instRomaENEA}
\affiliation{\instEri}
\affiliation{\instJuelich}
\affiliation{\instFuJen}
\affiliation{\instFudan}
\affiliation{\instGoettingen}
\affiliation{\instGifu}
\affiliation{\instSOKENDAI}
\affiliation{\instGyeongsang}
\affiliation{\instHanyang}
\affiliation{\instKEK}
\affiliation{\instJPARC}
\affiliation{\instHSE}
\affiliation{\instIISER}
\affiliation{\instIITBhubaneswar}
\affiliation{\instIITGuwahati}
\affiliation{\instIITHyderabad}
\affiliation{\instIITMadras}
\affiliation{\instIndiana}
\affiliation{\instIHEPRussia}
\affiliation{\instHEPHYVienna}
\affiliation{\instHiroshima}
\affiliation{\instIHEPChina}
\affiliation{\instIPP}
\affiliation{\instIOP}
\affiliation{\instIFIC}
\affiliation{\instFrascati}
\affiliation{\instNapoliINFN}
\affiliation{\instPadovaINFN}
\affiliation{\instPerugiaINFN}
\affiliation{\instPisaINFN}
\affiliation{\instRomaINFN}
\affiliation{\instRomaTreINFN}
\affiliation{\instTorinoINFN}
\affiliation{\instTriesteINFN}
\affiliation{\instJAEA}
\affiliation{\instMainz}
\affiliation{\instGiessen}
\affiliation{\instKarlsruhe}
\affiliation{\instISU}
\affiliation{\instKitasato}
\affiliation{\instKISTI}
\affiliation{\instKoreaUnivKU}
\affiliation{\instKSU}
\affiliation{\instKyungpook}
\affiliation{\instLPI}
\affiliation{\instLNNU}
\affiliation{\instLMU}
\affiliation{\instLuther}
\affiliation{\instMNITJaipur}
\affiliation{\instMPP}
\affiliation{\instMPGHLL}
\affiliation{\instMcGill}
\affiliation{\instMEPhI}
\affiliation{\instNagoya}
\affiliation{\instNagoyaIAR}
\affiliation{\instNagoyaKMI}
\affiliation{\instNaraWu}
\affiliation{\instNTUTaiwan}
\affiliation{\instNUUTaiwan}
\affiliation{\instKrakow}
\affiliation{\instNiigata}
\affiliation{\instNSU}
\affiliation{\instOkinawa}
\affiliation{\instOsakaCity}
\affiliation{\instRCNP}
\affiliation{\instPNNL}
\affiliation{\instPanjab}
\affiliation{\instPanjabPAU}
\affiliation{\instRIKENMSL}
\affiliation{\instSeoul}
\affiliation{\instSPU}
\affiliation{\instSoochow}
\affiliation{\instSoongsil}
\affiliation{\instLjubljanaJSI}
\affiliation{\instKyiv}
\affiliation{\instTata}
\affiliation{\instTUM}
\affiliation{\instTelAviv}
\affiliation{\instToho}
\affiliation{\instTohoku}
\affiliation{\instTitech}
\affiliation{\instTokyoMetropolitan}
\affiliation{\instUAS}
\affiliation{\instNapoliUNIV}
\affiliation{\instPadovaUNIV}
\affiliation{\instPerugiaUNIV}
\affiliation{\instPisaUNIV}
\affiliation{\instRomaTreUNIV}
\affiliation{\instTorinoUNIV}
\affiliation{\instTriesteUNIV}
\affiliation{\instMontreal}
\affiliation{\instIJCLab}
\affiliation{\instIPHC}
\affiliation{\instAdelaide}
\affiliation{\instBonn}
\affiliation{\instUBC}
\affiliation{\instCincinnati}
\affiliation{\instFlorida}
\affiliation{\instHawaii}
\affiliation{\instHeidelberg}
\affiliation{\instLjubljanaUniLJ}
\affiliation{\instLouisville}
\affiliation{\instMalaya}
\affiliation{\instLjubljanaUM}
\affiliation{\instMelbourne}
\affiliation{\instMississippi}
\affiliation{\instUOM}
\affiliation{\instPittsburgh}
\affiliation{\instUSTC}
\affiliation{\instSAlabama}
\affiliation{\instSCarolina}
\affiliation{\instSydney}
\affiliation{\instUTokyo}
\affiliation{\instIPMU}
\affiliation{\instVictoria}
\affiliation{\instVPI}
\affiliation{\instWayneState}
\affiliation{\instYamagata}
\affiliation{\instYerevan}
\affiliation{\instYonsei}
\affiliation{\instZZU}
  \author{F.~Abudin{\'e}n}\affiliation{\instTriesteINFN} 
  \author{I.~Adachi}\affiliation{\instKEK}\affiliation{\instSOKENDAI} 
  \author{R.~Adak}\affiliation{\instFudan} 
  \author{K.~Adamczyk}\affiliation{\instKrakow} 
  \author{P.~Ahlburg}\affiliation{\instBonn} 
  \author{J.~K.~Ahn}\affiliation{\instKoreaUnivKU} 
  \author{H.~Aihara}\affiliation{\instUTokyo} 
  \author{N.~Akopov}\affiliation{\instYerevan} 
  \author{A.~Aloisio}\affiliation{\instNapoliUNIV}\affiliation{\instNapoliINFN} 
  \author{F.~Ameli}\affiliation{\instRomaINFN} 
% ...

\collaboration{Belle II Collaboration}
 
\begin{abstract}
We present preliminary results on the reconstruction of the \bjpsikl decay, where
$\jpsi\to\mumu$ or \epem. Using a dataset corresponding to a luminosity
of $62.8\pm0.6\invfb$ collected by the Belle II
experiment at the SuperKEKB asymmetric energy \epem collider, we
measure a total of $267\pm21$ candidates with $\jpsi\to\mumu$ and
$226\pm20$ with with $\jpsi\to\epem$.
The quoted errors are statistical only.
\end{abstract}

\maketitle

{\renewcommand{\thefootnote}{\fnsymbol{footnote}}}
\setcounter{footnote}{0}

\tableofcontents

\clearpage

\section {Introduction and Motivation}
\label{sec:Introduction}
The Standard Model (SM) of electroweak interactions describes Charge-Parity (\CP) violation as a consequence of
an irreducible phase in the three-generation Cabibbo-Kobayashi-Maskawa (CKM) quark-mixing matrix.
In this framework, measurements of \CP asymmetries in the proper-time distribution of neutral $B$ decays to \CP
eigenstates containing a charmonium and \Kz meson provide a direct measurement of $\sin(2\phi_1)$,
where the angle $\phi_1$ of the Unitarity Triangle is given by
$\textrm{arg}({-{V_{cd}V_{cb}^\ast}/{V_{td}V_{tb}^\ast}})$ and
$V_{ij}$ are elements of the CKM matrix.

\CP violation in the \B meson system has been established by the \babar\,\cite{ref:babarS2b1}
and Belle\,\cite{ref:belleS2b1}
collaborations using a variety of such decays, among which the \bjpsikz one stands out for its clear
experimental signatures and clean theoretical interpretation, directly linking the measurement of the
time-dependent \CP asymmetry to the angle $\phi_1$.

The value of $\sin(2\phi_1)$ measured with \bjpsikz is basically free
of hadronic uncertainties and almost completely insensitive to the
effects of new physics, so it constitutes a reference point for the
SM\,\cite{ref:Kou_2019}.
Improving the precision of the measurement of $\sin(2\phi_1)$ is one of
the goals of the Belle II 
experimental program to verify the SM prediction that 
$\sin(2\phi_1)$ be equal in the tree-level \jpsi\Kz decays and in other
penguin-dominated decays described by the same CKM elements, with a
discrepancy being indicative of New Physics.

The \bjpsikl decay provides an independent measurement of
$\sin(2\phi_1)$ from \bjpsiks. In addition, the former decay having opposite \CP
eigenvalue of the latter, the decay \bjpsikl has a time-dependent \CP
asymmetry identical but opposite
in sign of that of \bjpsiks, providing an important check of
systematics effects on these measurements.

In this paper, we present preliminary results leading to the rediscovery
of the \bjpsikl decay at SuperKEKB using, for the first time in Belle
II, \KL mesons reconstructed from
hadronic neutral clusters in the \KL and muons (KLM) subdetector.
We identify signal events based on the variable \deltaE, representing
the difference between the reconstructed \B meson and the beam energy
in the center of mass.
In this preliminary result, the systematics uncertainties have not been
evaluated yet.

\section {The Belle II detector}
\label{sec:delle2_detector}
The Belle II detector is described in detail in Ref.\,\cite{Abe:2010gxa}. The detector
has a cylindrical structure around the beam pipe, placed partially
inside a solenoidal superconducting magnet providing a 1.5T magnetic
field. The innermost sub-detector is the vertex detector (VXD), formed
by two layers of silicon pixel sensors and four layers of silicon
strips, devoted to tracking and vertexing. It is surrounded by a large
central drift chamber (CDC), with small cells and filled with helium
ethane mixture, which provide precise measurement of momenta of
charged tracks as well as particle identification via energy loss
measurement (\dedx). Two Cherenkov detectors provide additional
particle identification: the Time of Propagation (TOP) counter in the
barrel region, and the Aerogel Ring Imaging Cherenkov (ARICH) in the
forward region. The final detector inside the solenoid is the
electromagnetic calorimeter (ECL), comprised of CsI(Tl) crystals,
dedicated to photon and electron identification and measurement. The
return yoke of the magnet is instrumented with scintillator strips and
resistive plate chambers, to provide identification of \KL mesons and
muons (KLM).
The coordinate system's axis coincides withe the solenoid axis, and is
roughly oriented with the electron beam. The polar angle $\vartheta$
is defined with respect the $z$ axis, and the azimuthal angle $\phi$
is measured in the transverse plane from the horizontal $x$ axis.

\section {Data and Monte Carlo samples}
\label{sec:data}
The dataset used for this analysis has been collected by Belle II in
2019 and 2020 at the SuperKEKB asymmetric energy \epem
collider\,\cite{ref:SuperKEKB}. The integrated luminosity collected
at a centre-of-mass (CM) energy corresponding to the \FourS resonance
is $62.8\pm0.6$\invfb, with an additional 9.2\invfb collected about 60 MeV
below the resonance (off-resonance dataset). This corresponds to
$(68.21\times10^6)\pm 0.09\%\mathrm{(stat)}\pm1.3\%\mathrm{(syst)}$
\BB pairs.\,\cite{ref:Bcounting}.

To study the selection efficiency, we used a sample of 40000
signal \bjpsimumukl and \bjpsieekl Monte Carlo (MC) events,
corresponding to 2.88\,\iab for each mode.
In addition, we used 40\,\ifb of generic (\smixed + \scharged + \sccbar)
events for background characterization. In the following, we refer to
these samples as ``\sampleA''  and ``\sampleB'', respectively.

\section {Event selection}
\label{sec:event_selection}
\subsection{\boldmath\jpsi selection}
\label{sec:jpsi_reco}
Signal candidate events are selected first by reconstructing 
\jpsi\to\mumu and \jpsi\to\epem, decays requiring the invariant mass of
the \jpsi candidate to be consistent with its PDG value.

We correct for final state radiation effects in $\jpsi\to\epem$ decays.
Due to the post-correction residual radiative tail, we also apply different
mass selections cuts ($m_{\mu\mu}\in [3.06,3.12]\,\gevcc$ and
$m_{ee}\in [3.04,3.12]\,\gevcc$ respectively) to the two \jpsi decay channels.
A tight particle identification requirement from the combined
information of all detectors is then applied to one of the \jpsi
tracks, and a looser requirement is applied to the second track.
Typical efficienciesl of the requirements for muons
are around 90\,\% and  95\,\% for tight and loose selection, respectively.
The corresponding values for electrons are 95\,\% and 97\,\%.

\subsection{\boldmath\KL selection}
The \KL mesons leave measurable signals in the ECL and KLM detectors.
In the present analysis, we only make use of neutral clusters in the KLM.

\begin {figure}[ht]
\begin{center}
  \includegraphics[ scale=0.48]{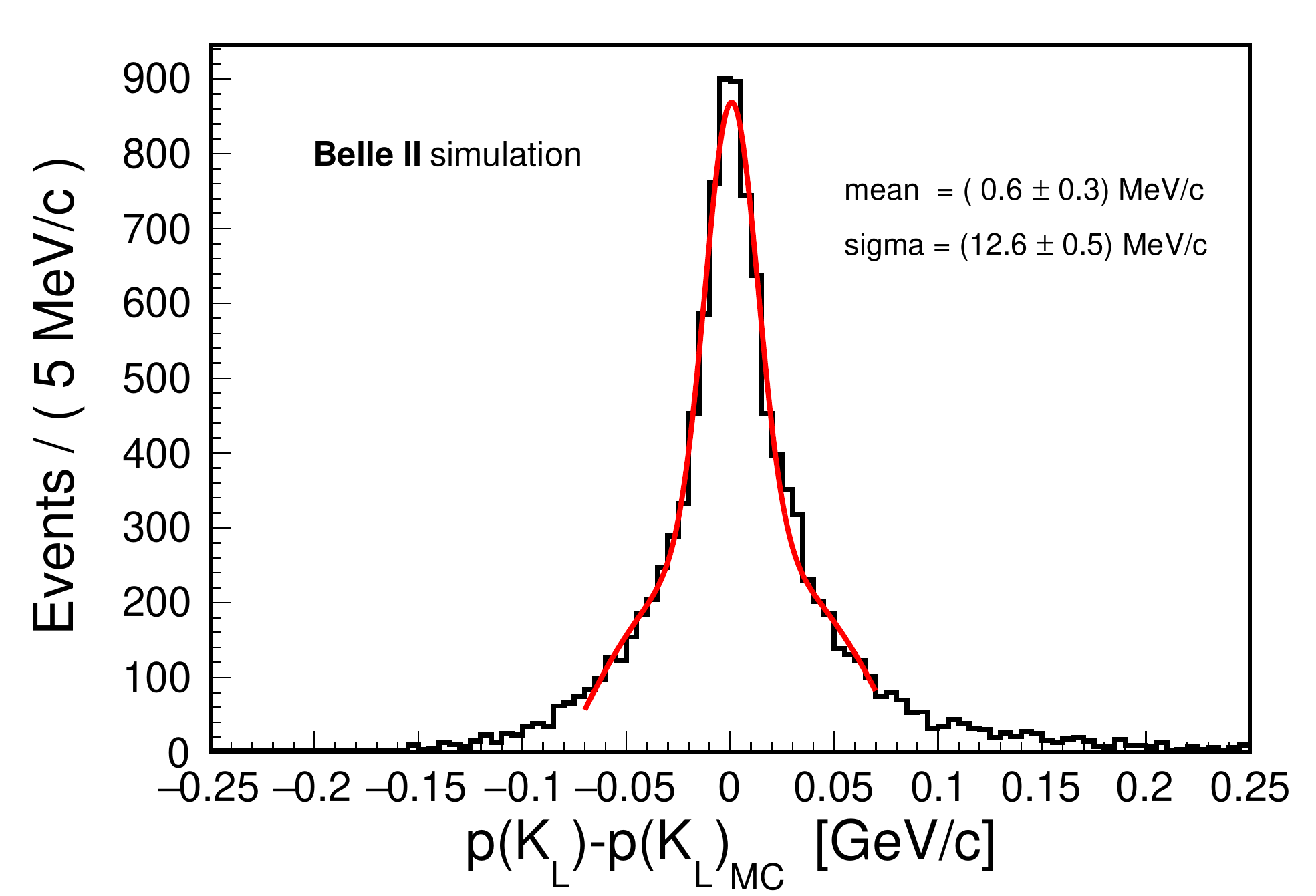}
\caption{Difference between calculated \KL
  momentum and  true \KL for truth-matched events in signal MC.}
\label{fig:kl_momentum}
\end{center}
\end{figure}

The energy deposition mechanism of \KL in the KLM does not allow
a usable measurement of the particle energy, but only of its direction
relative to the assumed origin at the interaction point.
We calculate the momentum of the \KL candidate from its
direction and the \jpsi reconstructed momentum, by requiring
momentum conservation and using the \Bz mass constraint in the
\bjpsikl decay.
The procedure correctly reconstructs the generated \KL momentum, with
a typical resolution of about 15\mevc, albeit with some non-gaussian
tails, as shown in \Fig{fig:kl_momentum}.
The \KL is reconstructed with an angular resolution of about 100\,mrad.

We enhance the purity of the KLM cluster using a multivariate boosted
decision tree (BDT) based on a number of KLM cluster shape and
topological variables and trained on signal \bjpsikl and generic
background simulated events.
The BDT output variable, \klongID,
is shown in \Fig{fig:klongID} for all
reconstructed \KL candidates and for those
truth-matched to a true \KL, respectively.
We use the requirement $\klongID\ge0.25$ in the following.
Further optimization of the BDT is in progress.

 \begin {figure}[ht]
 \begin{center} 
 \centering
 \includegraphics[width=0.6\textwidth]{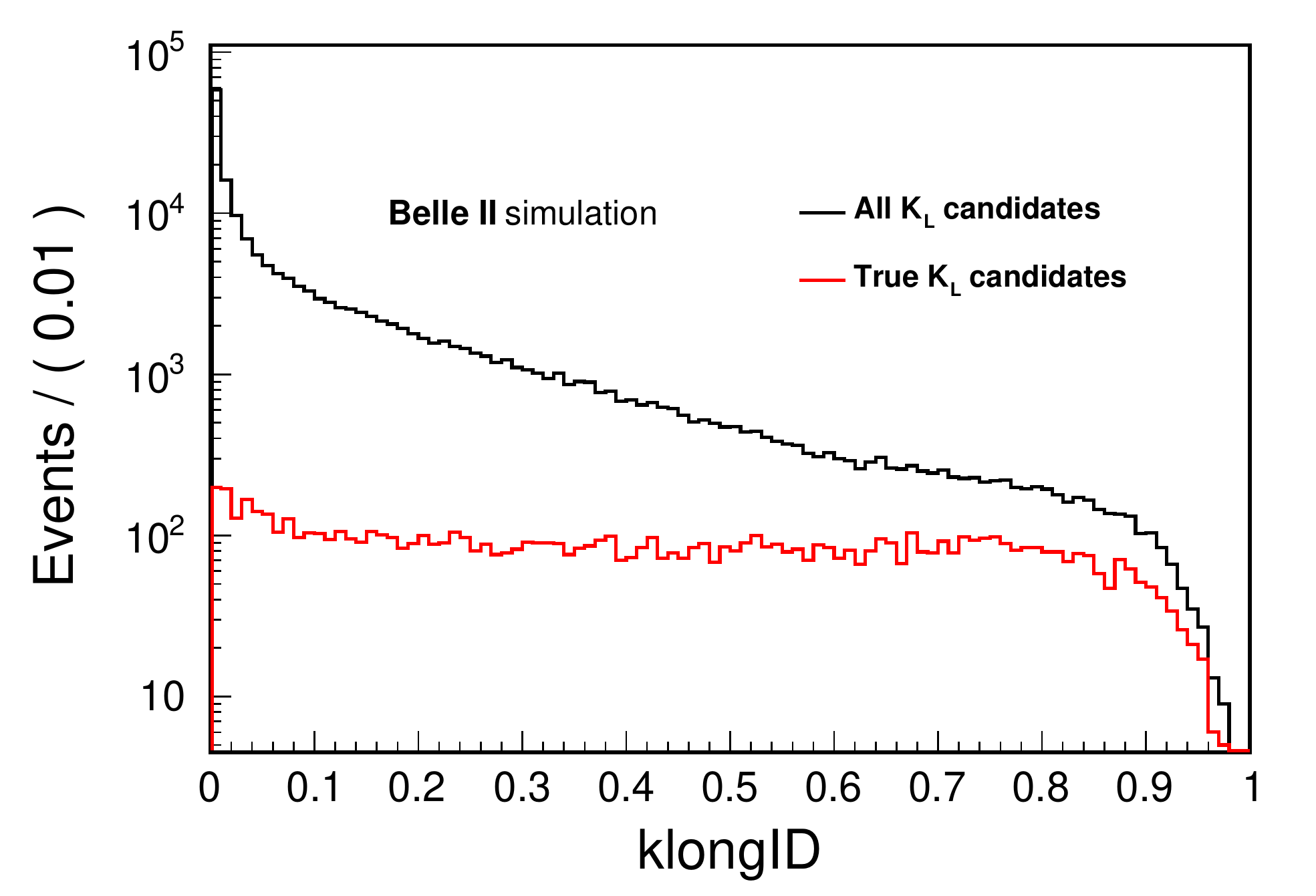}
 \caption{\klongID distribution of all \KL candidates (black curve)
   and truth-matched candidates (red curve) in the \sampleB.}
 \label{fig:klongID}
\end{center}
\end{figure}

Although the number of layers
in the KLM cluster is among the input BDT variables, we also
explicitely apply the requirement $\Nlayers\geqslant2$, which helps
eliminate spurious clusters.
\medskip

\section{\boldmath{\deltaE fits}}
Once a signal candidate has been selected we calculate the
difference between the \Bz candidate energy and the beam energy in the
CM frame, i.e., $\deltaE \equiv E_B ^* - E_\text{beam} ^*$, which for
genuine \bjpsikl is expected to be close to zero.

In a small number of cases (about 2\,\%), there are two
\bjpsikl candidates after the selection;
in these cases, we choose the \KL with the largest \klongID.
An alternative choice we have cosidered  is to select the candidate
with the smallest $|\deltaE|$.
In most cases the, two criteria select the
same candidate, which in the MC is also correctly matched to the
true \bjpsikl decay.
We conclude that the choice of the best candidate does not induce a bias
in the selection.

\subsection{PDF model for signal}
We model the signal \deltaE distribution with a Crystal Ball (CB)
probability density function (PDF)\,\cite{ref:CB_Oreglia,
  ref:CB_Skwarnicki}.
The signal PDF coefficients are determined by an unbinned maximum
likelihood (ML) fit to the \deltaE distribution in a high-statistics
MC sample of $\jpsi\KL$ events.
An example of such fits, after a selection requiring
$\klongID\geqslant0.25$,  is shown in \Fig{fig:deltae_signal_MC}.
Here and in the following, the fits are performed in the range
$\deltaE\in[-20,80]\MeV$.

\begin {figure}[ht]
\begin{center}
  \includegraphics[ scale=0.4]{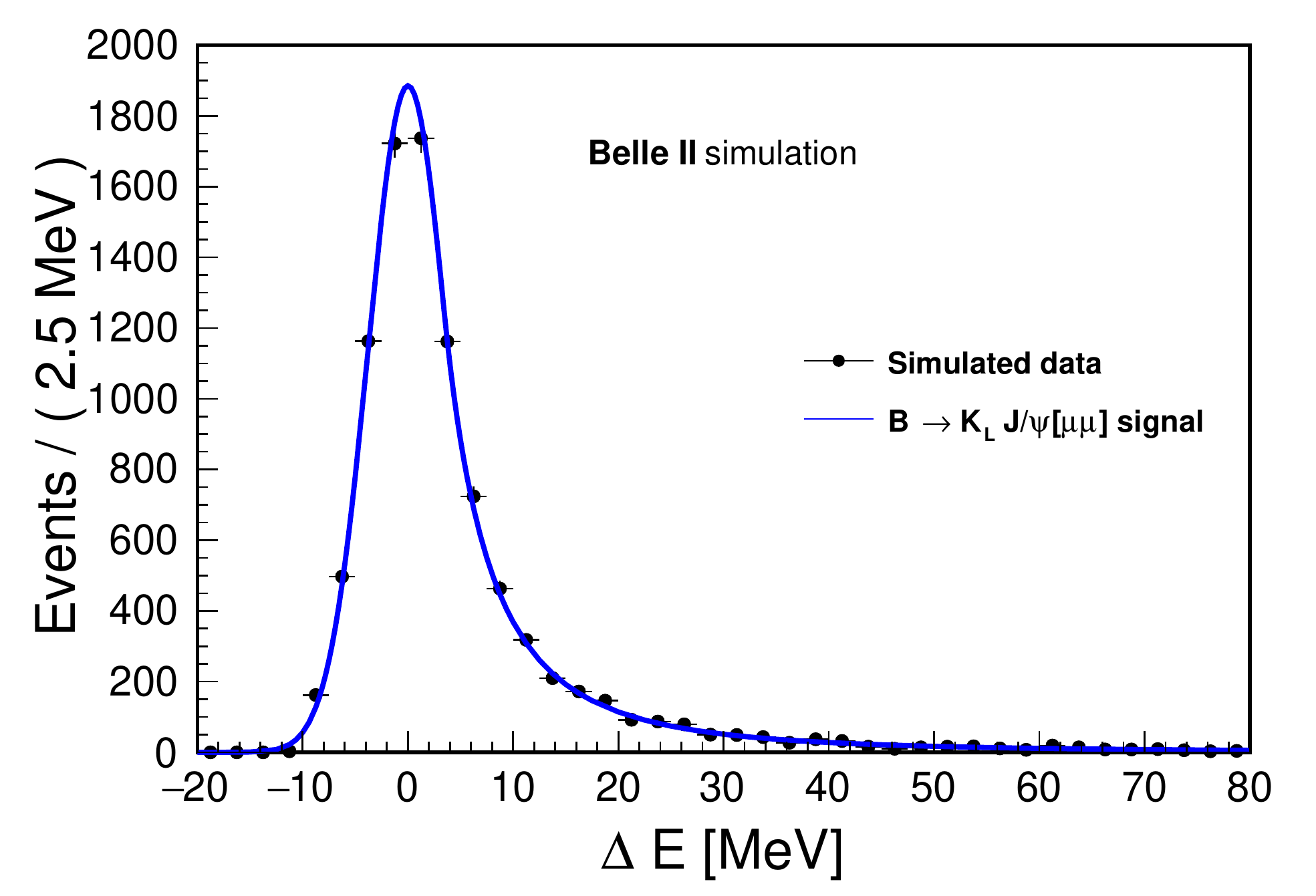}
  \includegraphics[ scale=0.4]{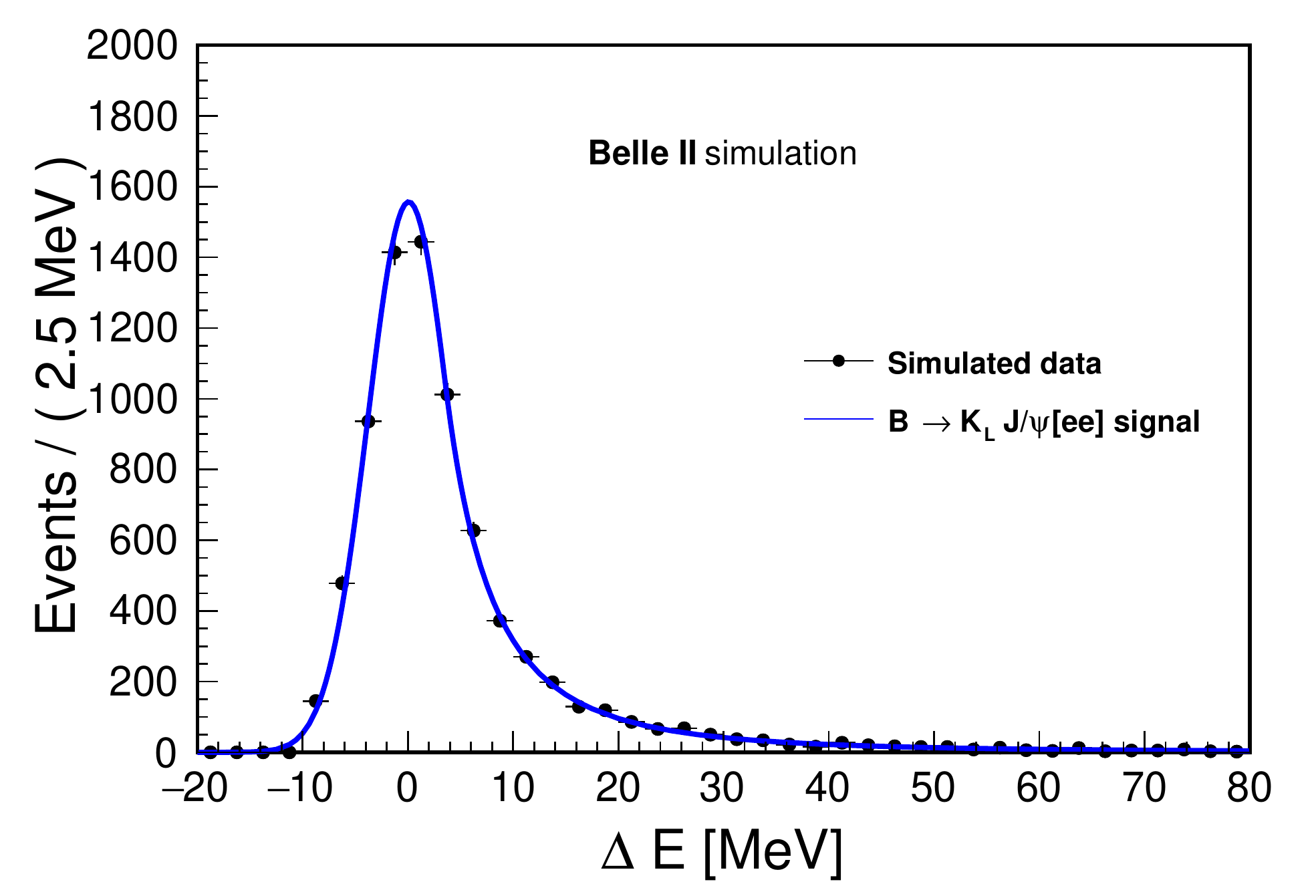}
  \caption{Signal \deltaE distribution of \bjpsikl candidate events for
  \jpsi\to\mumu final states (left)  and \jpsi\to\epem final states
  (right) from the signal MC sample.}
\label{fig:deltae_signal_MC}
\end{center}
\end{figure}

\subsection{Background description and parameterization}
\label{sec:backgrounds}
Background to \bjpsikl is essentially due to \BzBzb and \BpBm decays,
which according to MC simulation roughly contribute equally.
Thanks to the strong \jpsi signature, no events from the
$\epem\to c\bar{c}$ generic sample nor the off-resonance data survive
the analysis selection cuts.

We parameterize the background with an Argus
PDF\,\cite{ref:ALBRECHT1994217} for the combinatorial part and a
CB PDF to describe a possible peaking component, 
which according to the simulation could be expected mainly from
$\B\to\jpsi\Kstarz$ and $\B\to\jpsi\Kstarp$ decays.

The fraction of the peaking component in the background is determined
from fits to the \deltaE distributions of generic MC events (in which \bjpsikl
events are excluded from the \BzBzb sample).
We find the peaking background fraction
$\fpeak=(0.4\pm3.1)\%$ and $\fpeak=(0.0\pm3.1)\%$ in the
$\jpsi\to\mumu$ and  $\jpsi\to\epem$ final states, respectively.

In order to rely on MC simulations as little as possible, we estimate
the fake \jpsi and fake \KL backgrounds directly from data.
Wrongly reconstructed \jpsi candidates can be estimated using the
\jpsi mass sidebands, while fake {\KL} mesons are estimated using an
``\textit{anti-selection}'' of \KL clusters, with the requirements
$\Nlayers=1$, $\klongID<0.05$.

\section{Results}
\label{sec:data_results}
We determine the number of signal and background \bjpsikl events with
an unbinned ML fit to our $62.8\,\ifb$ dataset
in the \deltaE interval $[-20,+80]\mev$, as shown in
\Fig{fig:deltae_data_MORIOND2021}.
The background shape parameters determined in the background control
sample are fixed in the fit, as well as the peaking background
fraction estimated in the fit of the simulated background events.
The signal shape parameters are used as
starting values in the fit, but to minimize the dependence on
MC the sigma and mean of the CB are left free in the fit.
Finally, to extract the number of signal and background events, we also
float the relative normalization of the signal and background
distributions.
\begin {figure}[!ht]
\begin{center}
   \includegraphics[ scale=0.4]{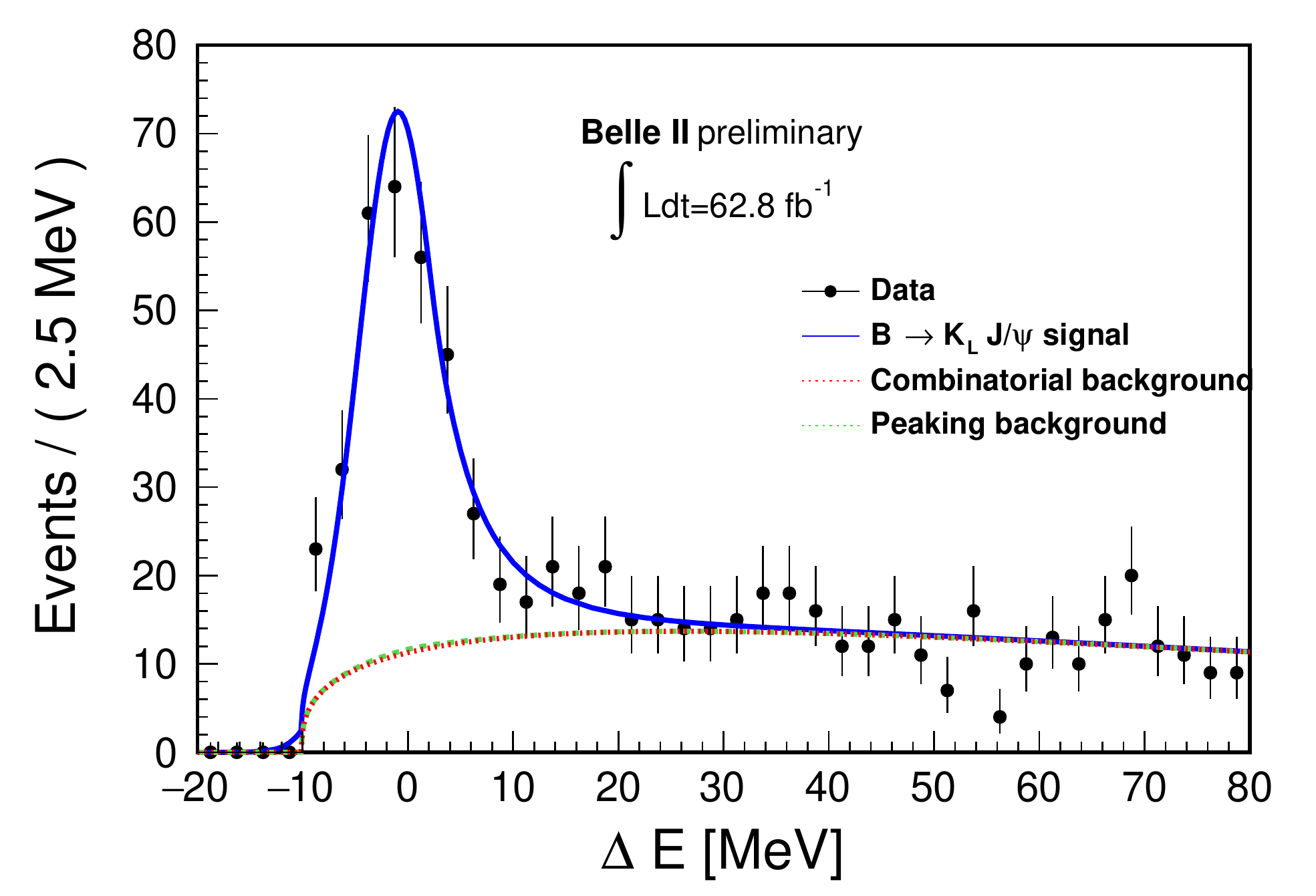}
   \includegraphics[ scale=0.4]{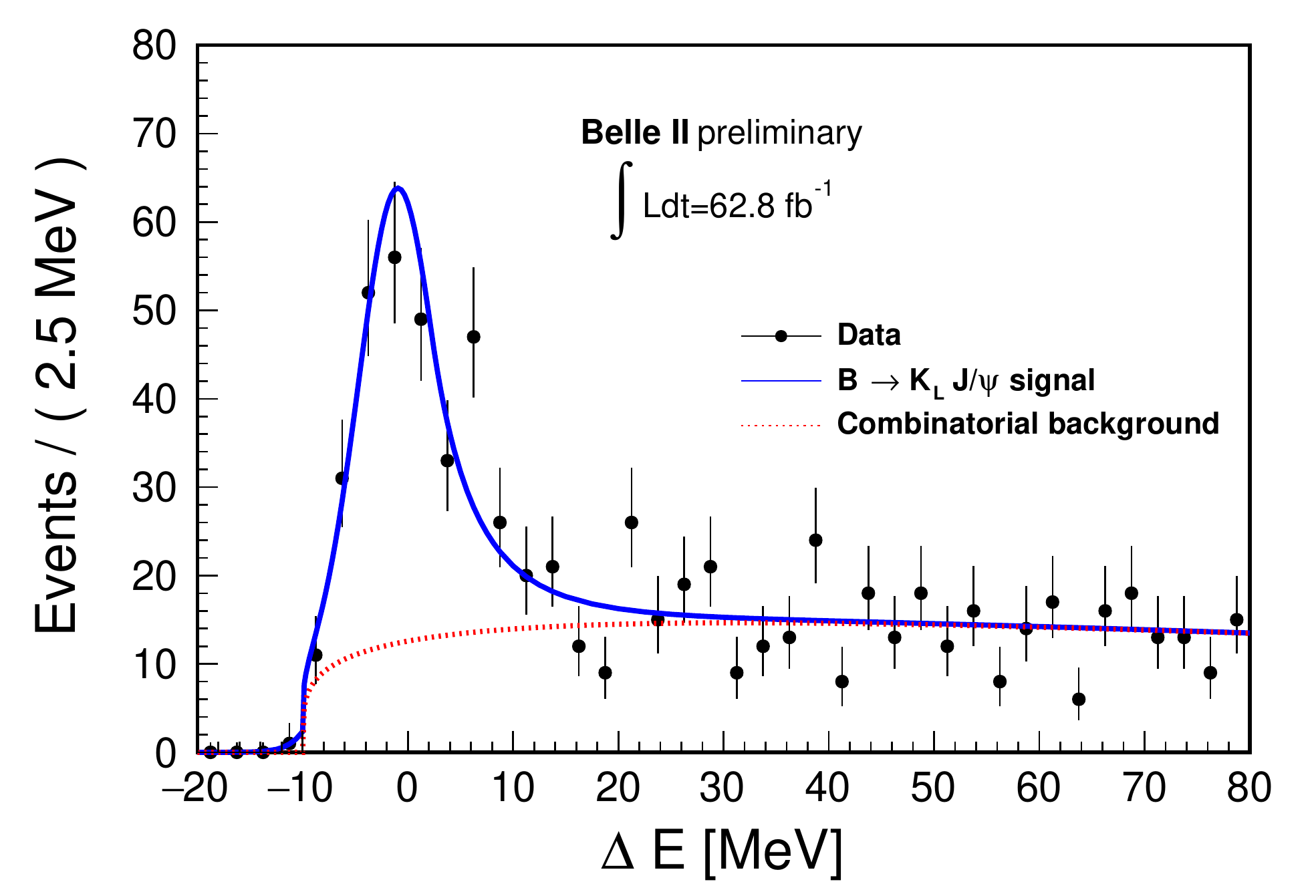}
  \caption{$\Delta E$ distribution of \bjpsikl candidate events for
    \jpsi\to\mumu final states (left)  and \jpsi\to\epem final states
    (right) for a dataset corresponding to a luminosity
    of $62.8\pm0.6\invfb$. The results of the unbinned ML fit are
    superimposed.}
  \label{fig:deltae_data_MORIOND2021}
\end{center}
\end{figure}
The results are $\Nsig(\mumu) = 267\pm21$, $\Nsig(\epem) = 226\pm20$.

A thorough evaluation of the systematics uncertainties has not been
performed yet. However, since we expect the one related to the
peaking background to be relatively large, we have done a conservative
estimate of this uncertainty by varying in the final fit the fraction
of peaking background (Sec.\ref{sec:backgrounds}) by twice
(95\,\% confidence interval) its statistical error.
This procedure yields $\Delta N_\mathrm{peaking}(\mumu)=28$, 
$\Delta N_\mathrm{peaking}(\epem)=31$.

\section{Conclusions}
In the present paper, we have shown preliminary results for the reconstruction
of the \bjpsikl decay in the first 62.8\,\ifb of integrated
luminosity collected by Belle II, which constitutes a rediscovery of
this decay.
The signal yields are:
\begin{table}[!ht]
  \begin{center}
    \renewcommand{\arraystretch}{1.1}
    \vspace*{0.2\baselineskip}
    \begin{tabular}{lrr}
      \Nsig(\mumu) &=& $267\pm21(\mathrm{stat})\pm28(\mathrm{peaking})~$  \\
      \Nsig(\epem)  &=& $226\pm20(\mathrm{stat})\pm31(\mathrm{peaking})$.
    \end{tabular}
    \label{tab:eventyields025}
  \end{center}
\end{table}

The overall signal yield obtained with this selection is consistent
with that observed by the Belle Collaboration, with similar purity.

Work is in progress to extend this study to include neutral clusters
reconstructed in the ECL, which will
significantly increase the signal sample.

Flavour tagging and tag and decay vertex time reconstruction will also
be added to allow the study of the time-dependent \CP violation and
the precise measurement of $\sin(2\phi_1)$.

\vspace{-0.5truecm}
\section*{Acknowledgements}
We thank the SuperKEKB group for the excellent operation of the
accelerator; the KEK cryogenics group for the efficient
operation of the solenoid; the KEK computer group for
on-site computing support; and the raw-data centers at
BNL, DESY, GridKa, IN2P3, and INFN for off-site computing support.
This work was supported by the following funding sources:
Science Committee of the Republic of Armenia Grant No. 20TTCG-1C010;
Australian Research Council and research grant Nos.
DP180102629, 
DP170102389, 
DP170102204, 
DP150103061, 
FT130100303, 
FT130100018,
and
FT120100745;
Austrian Federal Ministry of Education, Science and Research,
Austrian Science Fund No. P 31361-N36, and
Horizon 2020 ERC Starting Grant no. 947006 ``InterLeptons''; 
Natural Sciences and Engineering Research Council of Canada, Compute Canada and CANARIE;
Chinese Academy of Sciences and research grant No. QYZDJ-SSW-SLH011,
National Natural Science Foundation of China and research grant Nos.
11521505,
11575017,
11675166,
11761141009,
11705209,
and
11975076,
LiaoNing Revitalization Talents Program under contract No. XLYC1807135,
Shanghai Municipal Science and Technology Committee under contract No. 19ZR1403000,
Shanghai Pujiang Program under Grant No. 18PJ1401000,
and the CAS Center for Excellence in Particle Physics (CCEPP);
the Ministry of Education, Youth and Sports of the Czech Republic under Contract No.~LTT17020 and 
Charles University grants SVV 260448 and GAUK 404316;
European Research Council, 7th Framework PIEF-GA-2013-622527, 
Horizon 2020 ERC-Advanced Grants No. 267104 and 884719,
Horizon 2020 ERC-Consolidator Grant No. 819127,
Horizon 2020 Marie Sklodowska-Curie grant agreement No. 700525 `NIOBE,' 
and
Horizon 2020 Marie Sklodowska-Curie RISE project JENNIFER2 grant agreement No. 822070 (European grants);
L'Institut National de Physique Nucl\'{e}aire et de Physique des Particules (IN2P3) du CNRS (France);
BMBF, DFG, HGF, MPG, and AvH Foundation (Germany);
Department of Atomic Energy under Project Identification No. RTI 4002 and Department of Science and Technology (India);
Israel Science Foundation grant No. 2476/17,
United States-Israel Binational Science Foundation grant No. 2016113, and
Israel Ministry of Science grant No. 3-16543;
Istituto Nazionale di Fisica Nucleare and the research grants BELLE2;
Japan Society for the Promotion of Science,  Grant-in-Aid for Scientific Research grant Nos.
16H03968, 
16H03993, 
16H06492,
16K05323, 
17H01133, 
17H05405, 
18K03621, 
18H03710, 
18H05226,
19H00682, 
26220706,
and
26400255,
the National Institute of Informatics, and Science Information NETwork 5 (SINET5), 
and
the Ministry of Education, Culture, Sports, Science, and Technology (MEXT) of Japan;  
National Research Foundation (NRF) of Korea Grant Nos.
2016R1\-D1A1B\-01010135,
2016R1\-D1A1B\-02012900,
2018R1\-A2B\-3003643,
2018R1\-A6A1A\-06024970,
2018R1\-D1A1B\-07047294,
2019K1\-A3A7A\-09033840,
and
2019R1\-I1A3A\-01058933,
Radiation Science Research Institute,
Foreign Large-size Research Facility Application Supporting project,
the Global Science Experimental Data Hub Center of the Korea Institute of Science and Technology Information
and
KREONET/GLORIAD;
Universiti Malaya RU grant, Akademi Sains Malaysia and Ministry of Education Malaysia;
Frontiers of Science Program contracts
FOINS-296,
CB-221329,
CB-236394,
CB-254409,
and
CB-180023, and SEP-CINVESTAV research grant 237 (Mexico);
the Polish Ministry of Science and Higher Education and the National Science Center;
the Ministry of Science and Higher Education of the Russian Federation,
Agreement 14.W03.31.0026, and
the HSE University Basic Research Program, Moscow;
University of Tabuk research grants
S-0256-1438 and S-0280-1439 (Saudi Arabia);
Slovenian Research Agency and research grant Nos.
J1-9124
and
P1-0135; 
Agencia Estatal de Investigacion, Spain grant Nos.
FPA2014-55613-P
and
FPA2017-84445-P,
and
CIDEGENT/2018/020 of Generalitat Valenciana;
Ministry of Science and Technology and research grant Nos.
MOST106-2112-M-002-005-MY3
and
MOST107-2119-M-002-035-MY3, 
and the Ministry of Education (Taiwan);
Thailand Center of Excellence in Physics;
TUBITAK ULAKBIM (Turkey);
Ministry of Education and Science of Ukraine;
the US National Science Foundation and research grant Nos.
PHY-1807007 
and
PHY-1913789, 
and the US Department of Energy and research grant Nos.
DE-AC06-76RLO1830, 
DE-SC0007983, 
DE-SC0009824, 
DE-SC0009973, 
DE-SC0010073, 
DE-SC0010118, 
DE-SC0010504, 
DE-SC0011784, 
DE-SC0012704, 
DE-SC0021274; 
and
the Vietnam Academy of Science and Technology (VAST) under grant DL0000.05/21-23.

\bibliography{BELLE2-CONF-PH-2021-002.bib}
\end{document}